\documentclass{aa}
\usepackage{graphicx,natbib,amsmath,amssymb,fnpara}
\usepackage{txfonts}
\bibpunct[, ]{(}{)}{;}{a}{}{,}

\newcommand{\cxo}{\emph{Chandra}}
\newcommand{\xmm}{\emph{XMM-Newton}}

\begin{document}

   \title{Massive star-formation rates of $\gamma$-ray burst host galaxies: an unobscured view in X-rays}
   \titlerunning{X-ray star-formation rates of GRB host galaxies}

   \author{D.~Watson\inst{1}
           \and
           J.~Hjorth\inst{1}
           \and
           P.~Jakobsson\inst{1}
           \and
           K.~Pedersen\inst{1}
           \and
           S.~Patel\inst{2}
           \and
           C.~Kouveliotou\inst{3}
          }

   \offprints{D.~Watson, email: \texttt{darach@astro.ku.dk}}

   \institute{Niels Bohr Institute, Astronomical Observatory, University of Copenhagen, Juliane Maries Vej 30, DK-2100 Copenhagen \O, Denmark
              \and
              USRA (Universities Space Research Association), NSSTC, SD-50, 320 Sparkman Drive, Huntsville, AL 35805, USA
              \and
              NASA Marshall Space Flight Center, NSSTC, SD-50, 320 Sparkman Drive, Huntsville, AL 35805, USA}
   \date{Received / Accepted }

   \abstract{The hard X-ray (2--10\,keV) luminosity of a star-forming galaxy
             tracks its population of high mass X-ray binaries and is
             essentially unobscured. It is therefore a practically unbiased
             measure of star-formation in the host galaxies of $\gamma$-ray
             bursts (GRBs).  Using recent and archival observations of GRBs
             with the \xmm\ and \cxo\ X-ray observatories, limits are placed
             on the underlying X-ray emission from GRB hosts.  Useful limits
             on the current massive star-formation rates (SFRs), unaffected
             by obscuration, are obtained for the hosts of three low
             redshift GRBs: GRB\,980425, GRB\,030329 and GRB\,031203.  These
             limits show that though the specific SFRs may be high (as in
             dwarf starburst galaxies), none have massive obscured
             star-formation at the levels implied by the sub-mm detection of
             some GRB hosts. It is also shown that in cases where the faint
             luminosities of the late time afterglow or supernova emission
             are of interest, the contribution of the host galaxy to the
             X-ray flux may be significant.
   }

   \maketitle

%
%
\section{Introduction\label{introduction}}

Long-duration $\gamma$-ray bursts (GRBs) coincide with the demise of
certain massive stars
\citep[e.g.][]{2003Natur.423..847H,2003ApJ...591L..17S}. Because their
detection in $\gamma$-rays is unaffected by intervening gas and dust, they
provide a powerful, unbiased tracer of the location of high-redshift star-formation and thus allow a new means of identifying and studying distant
star-forming galaxies.
However, the majority of GRB host galaxies appear to be sub-luminous
($\langle R\rangle\sim 25$) and blue
\citep{1999ApJ...519L..13F,2003A&A...400..499L}.  Furthermore none are
Extremely Red Objects \citep[though][show that the host galaxy of GRB\,030115 is very red]{Levan:2004}
and few have detectable sub-mm
\citep{1999A&A...347...92S,2001A&A...380...81S,2003MNRAS.338....1B,%
2003ApJ...588...99B} or FIR fluxes
\citep{1999A&AS..138..459H,2000A&A...359..941H}.  GRBs also generally occur
within UV-bright parts of their hosts \citep{2002AJ....123.1111B}, which is
surprising if star-formation is generally obscured
\citep{2002A&A...384..848E,2002ApJ...568..651G,2003A&A...407..791M,2004AJ....127.1285S}. 
\citet{2003A&A...400..499L} have asked ``Are the hosts of gamma-ray bursts
sub-luminous and blue galaxies?'' This raises the further question of
whether a sizable proportion of global star-formation actually occurs in
small and relatively unobscured, modestly star-forming galaxies that are too
faint to appear in other surveys of star-formation activity, or whether GRBs
trace only a fraction of the star-forming population, for example due to
metallicity effects \citep{1999ApJ...524..262M,2003A&A...406L..63F}.

In this paper we examine the limits that can currently be placed on the
star-formation rates (SFRs) of GRB host galaxies using an unobscured, and
potentially unbiased tracer of star-formation: the high mass X-ray binary
(HMXB) population, via their aggregate hard X-ray (2--10\,keV) luminosity
\citep[see][]{2003MNRAS.339..793G,2003A&A...399...39R,2004MNRAS.347L..57G,%
2004A&A...419..849P}.  This measure has been shown to be linearly related
to the current massive SFR at redshifts up to $\sim1$. The following
relations are given by \citet{2003MNRAS.339..793G} and
\citet{2004A&A...419..849P} respectively for the massive and the total SFRs:
${\rm SFR}(\geq5\,M_{\sun}) = 1.5\times10^{-40}\times L_{2-10}\,~M_{\sun}\,\rm yr^{-1}$, and
${\rm SFR}(\geq0.1\,M_{\sun}) = 10^{-39}\times L^{\rm HMXB}_{2-10}\,~M_{\sun}\,\rm yr^{-1}$,
with luminosities in ergs\,s$^{-1}$. As long as the massive SFR is relatively high
\citep[$\gtrsim4.5\,M_{\sun}\,\rm yr^{-1}$;][]{2004MNRAS.347L..57G}, the
linear relation holds.  Where the SFR is lower than this, the relation
becomes non-linear due to the statistical properties of the combined
emission of a small number of discrete sources \citep{2004MNRAS.351.1365G};
${\rm SFR}(\geq5\,M_{\sun}) = (3.8\times10^{-40}\times L_{2-10})^{0.6}\,~M_{\sun}\,\rm yr^{-1}$ \citep{2003MNRAS.339..793G}.
In this paper, we derive upper limits to the SFR in GRB host galaxies by
assuming the 2--10\,keV luminosity is due exclusively to HMXBs \citep[taking
into account the 20\% scatter in the above relation;][]{2004A&A...419..849P}.

In Sect.~\ref{observations} selection criteria, details of the observations,
and our X-ray data reduction are described. The resulting detections and
limits are outlined in Sect.~\ref{results}.  The potential and limitations
of the technique are described in Sect.~\ref{X-ray-SFR} and the implications
of our results are discussed in Sect.~\ref{discussion}. 
A cosmology where $H_0=75$\,km\,s$^{-1}$\,Mpc$^{-1}$,
$\Omega_\Lambda = 0.7$ and $\Omega_{\rm m}=0.3$ is assumed throughout.

%
%
\section{Sample, observations, and data reduction\label{observations}}

All available follow-up observations of GRBs or X-Ray Flashes with
\cxo\ and \xmm\ performed before May 2004 were used in the analysis. Where
redshifts for the burst or its host galaxy were available in the literature,
these values were used (see Table~\ref{tab:all_obs}). Data
from earlier X-ray missions were not included. 

As an initial selection mechanism, results previously published or made
available in preliminary form (generally via GRB Coordinates Network
circulars), were examined in order to find datasets that might provide
strong constraints on the massive SFRs in GRB host galaxies.  Details of all
GRB localisations observed with \cxo\ and \xmm\ are provided in
Table~\ref{tab:all_obs}.  From these data it was immediately apparent that
only a small number provided potentially strong constraints: GRBs 980425,
030329 and 031203, all of which are associated with spectroscopically
confirmed SNe. The results for these hosts are produced separately in
Table~\ref{tab:best_obs}. 

It is not possible to disentangle the emission contributed by the late GRB
afterglow, the GRB (or supernova [SN]), or the HMXB and low mass X-ray
binary populations (or indeed a low-luminosity AGN) in any of these cases
except GRB\,980425.  However, we can at least derive useful upper limits on
the HMXB flux and therefore on the SFR in the host galaxy by considering the
aggregate flux from all of these components in a given galaxy.  Therefore it
is the total flux measurement that is considered below apart from the host
galaxy of GRB\,980425 (see Sect.~\ref{results}).

The \cxo\ datasets for GRB\,980425 and GRB\,031203 were reduced and analysed
using CIAO version 3.0.3 with the CALDB 2.26 version of the calibration
archive, and it is this analysis that is used in the paper.
To calibrate the count-rate to flux conversion, the count-rate was used as
the normalisation for an absorbed power-law model, with absorption set at
the Galactic value and the best-fit power-law photon index; where there was
insufficient data to determine the power-law index, $\Gamma=1.2$ was assumed
\citep[the emission in these star-forming galaxies is expected to be
dominated by the HMXB emission where the mean $\Gamma$ is
$\sim1.2$;][]{2004A&A...419..849P}.  For each burst the deepest limit is quoted.

\begin{table}
\caption{Data from the observations of all long-duration GRBs
         observed with \cxo\ or \xmm\ that place the lowest
         limit available on the flux.}
\label{tab:all_obs}
\setlength{\tabcolsep}{3pt}
\begin{minipage}{\columnwidth}
\renewcommand{\thempfootnote}{[\arabic{mpfootnote}]}
\begin{tabular}{@{}lcccccc@{}}
\hline\hline
GRB	& Obser-	&$t_{\rm burst}$& Range	& Flux					& $z$	& Luminosity\footnote{No bandpass or k-corrections applied}\\
	& vatory	& (days)	& (keV)	& \multicolumn{2}{l}{($10^{-15}$\,erg\,cm$^{-2}$\,s$^{-1}$)}	& ($10^{44}$\,erg\,s$^{-1}$)\\
\hline
040223  & N     &   0.75	&    2--10 &  $170$\footnote{\cite{2004GCN..2548....1T}} 	& --- 	& ---  \\
040106  & N     &   0.72	&    2--10 &  $400$\footnote{\cite{Reeves:2004}}		& ---   & ---  \\
031220  & C     &  19	&    2--10 &  $<3$\footnote{Assuming source \#1 \citep{2004GCN..2523....1G}} & ---	& --- \\
031203  & C   &  50	&    2--10 &  $4$\footnote{This paper ($\Gamma$=1.7)}	& 0.1055\footnote{\cite{2004ApJ...611..200P}}	& 0.099 \\
030723  & C     &  14	&   0.5--8 &  $<0.5$\footnote{\cite{2003GCN..2347....1B}}	& ---	& --- \\
030528  & C     &  12	&    2--10 &  $4.9$\footnote{\cite{2003GCN..2279....1B}}	& ---	& --- \\
030329  & N     &  258	&   0.5--2 &  $0.6$\footnote{\cite{2004A&A...423..861T}}	& 0.1685\footnote{\cite{2003Natur.423..847H}}	& 0.041 \\
030328  & C     &   0.64	&   0.5--3 &  $190$\footnote{\cite{2003GCN..2076....1B}}	& 1.52\footnote{\cite{2003GCN..1980....1M,2003GCN..1981....1R}}	& 2434 \\
030227  & N     &   0.81	&  0.2--10 &  $400$\footnote{\cite{2003ApJ...595L..29W}}	& 1.6$^[$\footnotemark[\value{mpfootnote}]$^]$	& 5818 \\
030226  & C     &   1.5	&    2--10 &  $32$\footnote{\cite{2003GCN..1924....1P}}		& 1.986\footnote{\cite{2003GCN..1894....1G}}	& 791 \\
021004  & C     &  52	&    2--10 &  $0.7$\footnote{\cite{2002GCN..1716....1S}}	& 2.33\footnote{\cite{2002A&A...396L..21M,2003ApJ...582L...5M,2003ApJ...588..387S}}	& 25.6 \\
020813  & C     &  43	&   0.6--6 &  $1450$\footnote{\cite{2002GCN..1504....1V}}	& 1.255\footnote{\cite{2003ApJ...584L..47B}}	& 11571 \\
020427  & C     &  17	&    2--10 &  $19$\footnote{\cite{2002GCN..1392....1F}}		& ---	& --- \\
020405  & C     &   1.7	&  0.2--10 &  $1360$\footnote{\cite{2003ApJ...587..128M}}	& 0.691\footnote{\cite{2003A&A...404..465M,2003ApJ...589..838P}}	& 2490 \\
020322  & N     &   0.86     	&  0.2--10 &  $320$\footnote{\cite{2002A&A...395L..41W}}	& ---	& --- \\
020321  & C     &   9.9	&  2--10  &  $<4$\footnote{see \cite{2002GCN..1348....1I}} & ---	& --- \\
020127  & C     &  14.3	&  2--10  &  $26$\footnote{\cite{2002GCN..1249....1F}}& ---	& --- \\
011211  & N     &   0.81	&  0.2--10 &  $80$\footnote{\cite{2002Natur.416..512R}}		& 2.140\footnote{\cite{2002AJ....124..639H}}	& 2375 \\
011130  & C     &  82	&  2-10  &  $<21$\footnote{No single source found, \citep{2002GCN..1272....1B}}  & ---	& --- \\
011030  & C     &  30	&   2--10   &  $7$ & ---	& --- \\
010222  & C     &  8.7	&    2--10 &  $72$\footnote{\cite{2001GCN..1023....1H}}		& 1.477\footnote{\cite{2001ApJ...554L.155J,2001A&A...374..382M,2003ApJ...587..135G}}	& 859 \\
010220  & N     &  0.87	&  0.2--10 &  $33$\footnote{\cite{2002A&A...393L...1W}}		& ---	& --- \\
001025A & N     &   910	&  0.3--12 &  $<4$\footnote{\cite{Pedersen:2004}}		& ---	& --- \\
000926  & C     &  13	&   1.5--8 &  $3.6$\footnote{\cite{2001ApJ...559..123H}}	& 2.0379\footnote{\cite{2003ApJ...586..128C}}	& 94.9 \\
000210  & C     &   0.93 	&    2--10 &  $180$\footnote{\cite{2002ApJ...577..680P}}	& 0.846$^[$\footnotemark[\value{mpfootnote}]$^]$	& 542 \\
991216  & C     &   1.6	&    2--10 &  $2300$\footnote{\cite{2000Sci...290..955P}}	& 1.02\footnote{\cite{1999GCN...496....1V,2000Sci...290..955P}}	& 10990 \\
980425  & C     &   1281	&    2--10 &  $13$\footnote{\cite{2004ApJ...608..872K}}	& 0.0085\footnote{\cite{1998Natur.395..670G}}	& 0.0018\\
\hline
\end{tabular}
\end{minipage}
\end{table}

\begin{table} 
\caption{X-ray observations of GRB positions with total SFR limits $<1000\,M_{\sun}$/yr.}
\label{tab:best_obs}
\setlength{\tabcolsep}{6pt}
\begin{minipage}{\columnwidth}
\begin{tabular}{@{}lcccc@{}}   
\hline\hline
GRB  &  Flux              & $z$  & Luminosity\footnote{Rest frame, 2--10\,keV}    & Equivalent Massive\\
\multicolumn{3}{c}{($10^{-15}$\,erg\,cm$^{-2}$\,s$^{-1}$)}    &($10^{41}$\,erg\,s$^{-1}$)     & SFR ($M_{\sun}$/yr)\\
\hline
031203  &  $6$\footnote{Assuming $\Gamma$=1.2}		& 0.1055\footnote{\cite{2004ApJ...611..200P}}	&   1.6  & 24\\
030329  &  $3$\footnote{\cite{2004A&A...423..861T}} 	& 0.1685\footnote{\cite{2003Natur.423..847H}}	&   2.1  & 32\\
980425  &  $7$\footnote{Afterglow contribution removed}	& 0.0085\footnote{\cite{1998Natur.395..670G}}	&  0.01  &  0.5\\
\hline
\hline
\end{tabular}
\end{minipage}
\end{table}

%
%

\section{Results\label{results}}

ESO\,184-G82, the host galaxy of GRB\,980425, is the closest known GRB host
and is the only one that is substantially spatially resolved with X-ray
instruments. The spiral optical morphology appears to exclude a galaxy
significantly larger than is seen in visible light. The X-ray flux within
the optical extent of the galaxy is entirely dominated by two point sources
$\sim1.5$\arcsec\ apart, one of which is coincident with the radio position
of SN1998bw and is almost certainly associated with it
\citep{2004ApJ...608..872K}.  Removing this source's contribution yields a
total 2--10\,keV flux for the galaxy of $7\pm3\times10^{-15}$\,
corresponding to a luminosity of $1\pm0.4\times10^{39}$\,erg\,s$^{-1}$, a
massive SFR of $0.5\pm0.3\,M_{\sun}$\,yr$^{-1}$ \citep[using the non-linear
relation for low-luminosity galaxies found by][]{2003MNRAS.339..793G} and a
total SFR of $2.8\pm1.9\,M_{\sun}$\,yr$^{-1}$ using the (Salpeter) IMF
adopted by \citet{2004A&A...419..849P} with the massive SFR derived
immediately above.  While it is noted that the galaxy appears to be actively
star-forming \citep{2000ApJ...542L..89F}, this is the first measure of its
global SFR to our knowledge. 

GRB\,030329 was monitored over 258 days with \xmm\ and in the final
observation, the afterglow was barely detected, with a 0.5--2.0\,keV flux of
$6.2\times10^{-16}$\,erg\,cm$^{-2}$\,s$^{-1}$ \citep{2004A&A...423..861T}.
The rest-frame 2--10\,keV luminosity was $\sim2\times10^{41}$\,erg\,s$^{-1}$, implying a
massive SFR of at most $31\pm13\,M_{\sun}$\,yr$^{-1}$ corresponding to a
total SFR of $\lesssim200\pm80\,M_{\sun}$\,yr$^{-1}$. Though the host is
very faint, estimates from optical observations (H$\alpha$ and [\ion{O}{ii}]
measures) suggest a total SFR of $\sim0.2\,M_{\sun}$\,yr$^{-1}$
\citep{2003Natur.423..847H} or $\sim0.5\,M_{\sun}$\,yr$^{-1}$
\citep{2003ApJ...599..394M}, consistent with the X-ray upper limit.
\citet{2003Natur.423..847H} suggest the host must be a dwarf starburst galaxy,
a finding confirmed by \citet{2003ApJ...599..394M}.

The host galaxy of GRB\,031203 (HG\,031203) was detected in the
near-infrared, however no optical/NIR GRB afterglow was discovered initially
\citep[though see][who subtracted the host galaxy light from early
data]{2004ApJ...609L...5M}. 
We recently obtained $\sim 30$ ks of Director's Discretionary Time to
observe the HG\,031203 with \cxo\ (Ramirez-Ruiz et al.\ in   
preparation). We found a faint X-ray point source with a flux of
$4\pm3\times10^{-15}$\,erg\,cm$^{-2}$\,s$^{-1}$ (2--10\,keV), assuming a
power-law photon index of 1.7, consistent with the extrapolation of the
afterglow decay rate observed in the previous two observations.  If we
instead assume a typically hard HMXB spectrum ($\Gamma=1.2$), this
corresponds to a luminosity of
$1.5\times10^{41}$\,erg\,s$^{-1}$ implying a massive SFR of at most
$24\pm17\,M_{\sun}$\,yr$^{-1}$, corresponding to a total SFR of
$\lesssim150\pm110\,M_{\sun}$\,yr$^{-1}$.  X-ray emission from the region
surrounding the galaxy out to 8\arcsec\ radius is consistent with the
background level. HG\,031203 is quite bright ($I\sim19.3$\,mag), though very
nearby ($z=0.1055$) for a GRB host galaxy, and is blue with low metallicity and
little internal extinction \citep{2004ApJ...611..200P}.  The total SFR is
$>11\,M_{\sun}$\,yr$^{-1}$ based on the H$\alpha$ luminosity
\citep{2004ApJ...611..200P}, consistent with the X-ray upper limit above.
Interestingly, this SFR implies a radio flux of $\sim0.3$\,mJy \citep[using
the assumptions and equation~1 of][]{2003ApJ...588...99B}, somewhat above an
upper limit published recently \citep{2004Natur.430..648S}. We infer
therefore, that a significant fraction of the flux detected during the
second observation derives from the host galaxy. The fact that a host galaxy
brighter than $\sim0.3$\,mJy is not observed, confirms the suggestion that
there is not a large obscured star-formation fraction in this galaxy.

It follows from the above analysis that a fraction of the X-ray flux
detected in observations of GRB afterglows comes from the HMXB
population of the host galaxy.  In the case of HG\,031203, based on the SFR
of $>11\,M_{\sun}$\,yr$^{-1}$, we expect at least one count in a 30\,ks
\cxo\ ACIS-S observation to have its origin in the HMXBs of the host galaxy,
rather than belonging to the GRB/SN.  This is about 10\% of the detected
counts in the most recent observation.

%
%

\section{X-ray luminosity as a SFR indicator\label{X-ray-SFR}}
The hard X-ray (2--10\,keV) SFR indicator is especially useful in cases of
extreme obscuration as the X-rays are only seriously attenuated at low
($<2$\,keV) energies, unless the absorbing column density is
$\gtrsim10^{23}$\,cm$^{-2}$, corresponding to $A_{V}\gtrsim100$ throughout
most of the starburst (at the gas-to-dust ratios of the Galaxy). Another key
advantage of the technique is the relatively high spatial resolution
($\sim0.5$\arcsec\ with \cxo) and high localisation accuracy (typically
$\sim0.3$\arcsec\ with \cxo\ and $\sim0.8$\arcsec\ with \xmm) that means
there is essentially no risk of misidentification of the source.  In some
cases there is the possibility of localising the dominant source of
star-formation within a galaxy. This is in contrast to other observing bands
that are unaffected by dust obscuration, the FIR, sub-mm and radio
wavelengths, where the beam size is often $\gtrsim10$\arcsec.  The
possibility of localising the sources of star-formation within a galaxy
without having to worry about obscuration is especially interesting as GRBs
generally occur within UV-bright parts of their hosts
\citep{2002AJ....123.1111B}.  Finally, it is likely that the HMXB population
traces the SFR in the same way as GRBs, since both are affected by many of
the same effects (short-lived massive stars, possibly metallicity, binarity
etc.).

The principle limitation of this estimate is the lack of sensitivity at
high redshift, since even a moderately long (100\,ks) \cxo\ exposure of a
GRB host can only give a massive SFR limit of
$\sim500\,M_{\sun}$\,yr$^{-1}$ at redshifts $z\sim1$.  It should be
mentioned however, that the linear relation between exposure time and
limiting depth using \cxo\ is favourable to making very long observations of
at least a few sources, in particular those at low redshift. 

Even the small sample with useful constraints examined in this paper is
likely to be biased to some extent. In the first case we obtain useful
limits only at relatively low redshift, while it is apparent that the space
density of ultraluminous infrared galaxies (ULIRGs) increases dramatically from low redshift to redshift
$\sim1$ \citep{2002A&A...384..848E}.  Second, it may be expected that
fainter (and possibly X-ray rich) GRBs dominate the observed GRB rate at low
$z$, a factor potentially related to many parameters (metallicity,
orientation etc.). Otherwise, in terms of obscuration
effects, this small sample should be unaffected.

\section{SFRs of GRB host galaxies\label{discussion}}

A few GRB host galaxies have detectable FIR, sub-mm and/or radio detections,
implying that they are ULIRGs:
GRB\,970508, GRB\,000418, GRB\,000210, GRB\,980703 and GRB\,010222
\citep{2000A&A...359..941H,2001A&A...380...81S,2003ApJ...588...99B,2004MNRAS.352.1073T}.
It has been suggested that a considerable fraction of GRBs are hosted in
ULIRGs \citep[$\sim20\%$,][]{2003ApJ...588...99B}. 
Though the large beam-size for the FIR, sub-mm and radio observations make
an unambiguous association between the GRB and the ULIRG somewhat uncertain,
the probability of a chance association is low.  Curiously, these massive
star-forming galaxies which should have fairly high internal extinctions
exhibit blue colours and very low extinction in optical and NIR observations
\citep{2002ApJ...565..829F,2003A&A...400..127G,2003A&A...409..123G,2003A&A...400..499L},
making them appear at these wavelengths to be dust poor, star-forming dwarf
galaxies. 
From the X-ray limits presented here however, it is
apparent that the hosts of GRB\,980425, GRB\,030329, and GRB\,031203 are
unlike the host galaxies of the sub-mm--detected galaxies 
in spite of the similarities in their optical/NIR properties (blue colours, apparently moderate SFR).

Although a large, deep X-ray sample of GRB hosts would be very valuable, it
would be expensive in terms of observing time.  The association of ULIRGs as
hosts of some GRBs can be tested directly however with only a few
observations; long exposures with \cxo\ of the GRB ULIRG host galaxies, at
least in the cases of the two strong claims for a ULIRG connection with
lower redshifts, GRB\,000210 and GRB\,000418 would decide the issue. \cxo\
imaging in hard X-rays will allow an unambiguous association to be made and
could confirm the sub-mm detection in an exposure time of $\sim250$\,ks.

Finally, it should be noted that in observations where the very faint fluxes
associated with the late time afterglow or SN are of interest, the flux
contribution from the host galaxy may be significant and should be accounted
for.

\small
\bibliography{mnemonic,grbs}

\begin{thebibliography}{73}
\expandafter\ifx\csname natexlab\endcsname\relax\def\natexlab#1{#1}\fi

\bibitem[{{Barnard} {et~al.}(2003){Barnard}, {Blain}, {Tanvir}, {Natarajan},
  {Smith}, {Wijers}, {Kouveliotou}, {Rol}, {Tilanus}, \&
  {Vreeswijk}}]{2003MNRAS.338....1B}
{Barnard}, V.~E., {Blain}, A.~W., {Tanvir}, N.~R., {et~al.} 2003, MNRAS, 338, 1

\bibitem[{{Barth} {et~al.}(2003){Barth}, {Sari}, {Cohen}, {Goodrich}, {Price},
  {Fox}, {Bloom}, {Soderberg}, \& {Kulkarni}}]{2003ApJ...584L..47B}
{Barth}, A.~J., {Sari}, R., {Cohen}, M.~H., {et~al.} 2003, ApJ, 584, L47

\bibitem[{{Berger} {et~al.}(2003){Berger}, {Cowie}, {Kulkarni}, {Frail},
  {Aussel}, \& {Barger}}]{2003ApJ...588...99B}
{Berger}, E., {Cowie}, L.~L., {Kulkarni}, S.~R., {et~al.} 2003, ApJ, 588, 99

\bibitem[{{Bloom} {et~al.}(2002){Bloom}, {Kulkarni}, \&
  {Djorgovski}}]{2002AJ....123.1111B}
{Bloom}, J.~S., {Kulkarni}, S.~R., \& {Djorgovski}, S.~G. 2002, AJ, 123, 1111

\bibitem[{{Butler} {et~al.}(2003{\natexlab{a}}){Butler}, {Dullighan}, {Ford},
  {Ricker}, {Vanderspek}, {Hurley}, {Jernigan}, \&
  {Lamb}}]{2003GCN..2279....1B}
{Butler}, N., {Dullighan}, A., {Ford}, P., {et~al.} 2003{\natexlab{a}}, GCN
  Circ., 2279

\bibitem[{{Butler} {et~al.}(2003{\natexlab{b}}){Butler}, {Ford}, {Ricker},
  {Vanderspek}, {Villasenor}, {Lamb}, {Garmire}, {Piro}, \&
  {Jernigan}}]{2003GCN..2347....1B}
{Butler}, N., {Ford}, P., {Ricker}, G., {et~al.} 2003{\natexlab{b}}, GCN Circ.,
  2347

\bibitem[{{Butler} {et~al.}(2003{\natexlab{c}}){Butler}, {Marshall}, {Ford},
  {Vanderspek}, {Ricker}, {Jernigan}, \& {Lamb}}]{2003GCN..2076....1B}
{Butler}, N., {Marshall}, H., {Ford}, P., {et~al.} 2003{\natexlab{c}}, GCN
  Circ., 2076

\bibitem[{{Butler} {et~al.}(2002){Butler}, {Monnelly}, {Ricker}, {Doty},
  {Ford}, {Vanderspek}, {Crew}, {Dullighan}, {Lamb}, \&
  {Plucinsky}}]{2002GCN..1272....1B}
{Butler}, N., {Monnelly}, G., {Ricker}, G., {et~al.} 2002, GCN Circ., 1272

\bibitem[{{Castro} {et~al.}(2003){Castro}, {Galama}, {Harrison}, {Holtzman},
  {Bloom}, {Djorgovski}, \& {Kulkarni}}]{2003ApJ...586..128C}
{Castro}, S., {Galama}, T.~J., {Harrison}, F.~A., {et~al.} 2003, ApJ, 586, 128

\bibitem[{{Elbaz} {et~al.}(2002){Elbaz}, {Cesarsky}, {Chanial}, {Aussel},
  {Franceschini}, {Fadda}, \& {Chary}}]{2002A&A...384..848E}
{Elbaz}, D., {Cesarsky}, C.~J., {Chanial}, P., {et~al.} 2002, A\&A, 384, 848

\bibitem[{{Fox}(2002{\natexlab{a}})}]{2002GCN..1249....1F}
{Fox}, D.~W. 2002{\natexlab{a}}, GCN Circ., 1249

\bibitem[{{Fox}(2002{\natexlab{b}})}]{2002GCN..1392....1F}
{Fox}, D.~W. 2002{\natexlab{b}}, GCN Circ., 1392

\bibitem[{{Frail} {et~al.}(2002){Frail}, {Bertoldi}, {Moriarty-Schieven},
  {Berger}, {Price}, {Bloom}, {Sari}, {Kulkarni}, {Gerardy}, {Reichart},
  {Djorgovski}, {Galama}, {Harrison}, {Walter}, {Shepherd}, {Halpern}, {Peck},
  {Menten}, {Yost}, \& {Fox}}]{2002ApJ...565..829F}
{Frail}, D.~A., {Bertoldi}, F., {Moriarty-Schieven}, G.~H., {et~al.} 2002, ApJ,
  565, 829

\bibitem[{{Fruchter} {et~al.}(1999){Fruchter}, {Thorsett}, {Metzger}, {Sahu},
  {Petro}, {Livio}, {Ferguson}, {Pian}, {Hogg}, {Galama}, {Gull},
  {Kouveliotou}, {Macchetto}, {van Paradijs}, {Pedersen}, \&
  {Smette}}]{1999ApJ...519L..13F}
{Fruchter}, A.~S., {Thorsett}, S.~E., {Metzger}, M.~R., {et~al.} 1999, ApJ,
  519, L13

\bibitem[{{Fynbo} {et~al.}(2000){Fynbo}, {Holland}, {Andersen}, {Thomsen},
  {Hjorth}, {Bj{\" o}rnsson}, {Jaunsen}, {Natarajan}, \&
  {Tanvir}}]{2000ApJ...542L..89F}
{Fynbo}, J.~P.~U., {Holland}, S., {Andersen}, M.~I., {et~al.} 2000, ApJ, 542,
  L89

\bibitem[{{Fynbo} {et~al.}(2003){Fynbo}, {Jakobsson}, {M{\o}ller}, {Hjorth},
  {Thomsen}, {Andersen}, {Fruchter}, {Gorosabel}, {Holland}, {Ledoux},
  {Pedersen}, {Rhoads}, {Weidinger}, \& {Wijers}}]{2003A&A...406L..63F}
{Fynbo}, J.~P.~U., {Jakobsson}, P., {M{\o}ller}, P., {et~al.} 2003, A\&A, 406,
  L63

\bibitem[{{Galama} {et~al.}(2003){Galama}, {Reichart}, {Brown}, {Kimble},
  {Price}, {Berger}, {Frail}, {Kulkarni}, {Yost}, {Gal-Yam}, {Bloom},
  {Harrison}, {Sari}, {Fox}, \& {Djorgovski}}]{2003ApJ...587..135G}
{Galama}, T.~J., {Reichart}, D., {Brown}, T.~M., {et~al.} 2003, ApJ, 587, 135

\bibitem[{{Galama} {et~al.}(1998){Galama}, {Vreeswijk}, {van Paradijs},
  {Kouveliotou}, {Augusteijn}, {Bohnhardt}, {Brewer}, {Doublier}, {Gonzalez},
  {Leibundgut}, {Lidman}, {Hainaut}, {Patat}, {Heise}, {in 't Zand}, {Hurley},
  {Groot}, {Strom}, {Mazzali}, {Iwamoto}, {Nomoto}, {Umeda}, {Nakamura},
  {Young}, {Suzuki}, {Shigeyama}, {Koshut}, {Kippen}, {Robinson}, {de Wildt},
  {Wijers}, {Tanvir}, {Greiner}, {Pian}, {Palazzi}, {Frontera}, {Masetti},
  {Nicastro}, {Feroci}, {Costa}, {Piro}, {Peterson}, {Tinney}, {Boyle},
  {Cannon}, {Stathakis}, {Sadler}, {Begam}, \& {Ianna}}]{1998Natur.395..670G}
{Galama}, T.~J., {Vreeswijk}, P.~M., {van Paradijs}, J., {et~al.} 1998, Nat,
  395, 670

\bibitem[{{Gendre} {et~al.}(2004){Gendre}, {de Pasquale}, {Piro}, {Costa},
  {Feroci}, {Garcia}, {Antonelli}, {Garmire}, {Ricker}, {Tagliaferri}, \& {in't
  Zand}}]{2004GCN..2523....1G}
{Gendre}, B., {de Pasquale}, M., {Piro}, L., {et~al.} 2004, GCN Circ., 2523

\bibitem[{{Gilfanov} {et~al.}(2004{\natexlab{a}}){Gilfanov}, {Grimm}, \&
  {Sunyaev}}]{2004MNRAS.347L..57G}
{Gilfanov}, M., {Grimm}, H.-J., \& {Sunyaev}, R. 2004{\natexlab{a}}, MNRAS,
  347, L57

\bibitem[{{Gilfanov} {et~al.}(2004{\natexlab{b}}){Gilfanov}, {Grimm}, \&
  {Sunyaev}}]{2004MNRAS.351.1365G}
{Gilfanov}, M., {Grimm}, H.-J., \& {Sunyaev}, R. 2004{\natexlab{b}}, MNRAS,
  351, 1365

\bibitem[{{Goldader} {et~al.}(2002){Goldader}, {Meurer}, {Heckman}, {Seibert},
  {Sanders}, {Calzetti}, \& {Steidel}}]{2002ApJ...568..651G}
{Goldader}, J.~D., {Meurer}, G., {Heckman}, T.~M., {et~al.} 2002, ApJ, 568, 651

\bibitem[{{Gorosabel} {et~al.}(2003{\natexlab{a}}){Gorosabel}, {Christensen},
  {Hjorth}, {Fynbo}, {Pedersen}, {Jensen}, {Andersen}, {Lund}, {Jaunsen},
  {Castro Cer{\' o}n}, {Castro-Tirado}, {Fruchter}, {Greiner}, {Pian},
  {Vreeswijk}, {Burud}, {Frontera}, {Kaper}, {Klose}, {Kouveliotou}, {Masetti},
  {Palazzi}, {Rhoads}, {Rol}, {Salamanca}, {Tanvir}, {Wijers}, \& {van den
  Heuvel}}]{2003A&A...400..127G}
{Gorosabel}, J., {Christensen}, L., {Hjorth}, J., {et~al.} 2003{\natexlab{a}},
  A\&A, 400, 127

\bibitem[{{Gorosabel} {et~al.}(2003{\natexlab{b}}){Gorosabel}, {Klose},
  {Christensen}, {Fynbo}, {Hjorth}, {Greiner}, {Tanvir}, {Jensen}, {Pedersen},
  {Holland}, {Lund}, {Jaunsen}, {Castro Cer{\' o}n}, {Castro-Tirado},
  {Fruchter}, {Pian}, {Vreeswijk}, {Burud}, {Frontera}, {Kaper}, {Kouveliotou},
  {Masetti}, {Palazzi}, {Rhoads}, {Rol}, {Salamanca}, {Wijers}, \& {van den
  Heuvel}}]{2003A&A...409..123G}
{Gorosabel}, J., {Klose}, S., {Christensen}, L., {et~al.} 2003{\natexlab{b}},
  A\&A, 409, 123

\bibitem[{{Greiner} {et~al.}(2003){Greiner}, {Ries}, {Barwig}, {Fynbo}, \&
  {Klose}}]{2003GCN..1894....1G}
{Greiner}, J., {Ries}, C., {Barwig}, H., {Fynbo}, J., \& {Klose}, S. 2003, GCN
  Circ., 1894

\bibitem[{{Grimm} {et~al.}(2003){Grimm}, {Gilfanov}, \&
  {Sunyaev}}]{2003MNRAS.339..793G}
{Grimm}, H.-J., {Gilfanov}, M., \& {Sunyaev}, R. 2003, MNRAS, 339, 793

\bibitem[{{Hanlon} {et~al.}(2000){Hanlon}, {Laureijs}, {Metcalfe}, {McBreen},
  {Altieri}, {Castro-Tirado}, {Claret}, {Costa}, {Delaney}, {Feroci},
  {Frontera}, {Galama}, {Gorosabel}, {Groot}, {Heise}, {Kessler},
  {Kouveliotou}, {Palazzi}, {van Paradijs}, {Piro}, \&
  {Smith}}]{2000A&A...359..941H}
{Hanlon}, L., {Laureijs}, R.~J., {Metcalfe}, L., {et~al.} 2000, A\&A, 359, 941

\bibitem[{{Hanlon} {et~al.}(1999){Hanlon}, {Metcalfe}, {Delaney}, {Laureijs},
  {McBreen}, {Smith}, {Altieri}, {Castro-Tirado}, {Costa}, {Feroci},
  {Frontera}, {Galama}, {Gorosabel}, {Groot}, {Heise}, {Kouveliotou},
  {Palazzi}, {van Paradijs}, {Piro}, \& {Kessler}}]{1999A&AS..138..459H}
{Hanlon}, L., {Metcalfe}, L., {Delaney}, M., {et~al.} 1999, A\&AS, 138, 459

\bibitem[{{Harrison} {et~al.}(2001{\natexlab{a}}){Harrison}, {Yost}, \&
  {Kulkarni}}]{2001GCN..1023....1H}
{Harrison}, F.~A., {Yost}, S.~A., \& {Kulkarni}, S.~R. 2001{\natexlab{a}}, GCN
  Circ., 1023

\bibitem[{{Harrison} {et~al.}(2001{\natexlab{b}}){Harrison}, {Yost}, {Sari},
  {Berger}, {Galama}, {Holtzman}, {Axelrod}, {Bloom}, {Chevalier}, {Costa},
  {Diercks}, {Djorgovski}, {Frail}, {Frontera}, {Hurley}, {Kulkarni},
  {McCarthy}, {Piro}, {Pooley}, {Price}, {Reichart}, {Ricker}, {Shepherd},
  {Schmidt}, {Walter}, \& {Wheeler}}]{2001ApJ...559..123H}
{Harrison}, F.~A., {Yost}, S.~A., {Sari}, R., {et~al.} 2001{\natexlab{b}}, ApJ,
  559, 123

\bibitem[{{Hjorth} {et~al.}(2003){Hjorth}, {Sollerman}, {M{\o}ller}, {Fynbo},
  {Woosley}, {Kouveliotou}, {Tanvir}, {Greiner}, {Andersen}, {Castro-Tirado},
  {Castro Cer{\' o}n}, {Fruchter}, {Gorosabel}, {Jakobsson}, {Kaper}, {Klose},
  {Masetti}, {Pedersen}, {Pedersen}, {Pian}, {Palazzi}, {Rhoads}, {Rol}, {van
  den Heuvel}, {Vreeswijk}, {Watson}, \& {Wijers}}]{2003Natur.423..847H}
{Hjorth}, J., {Sollerman}, J., {M{\o}ller}, P., {et~al.} 2003, Nat, 423, 847

\bibitem[{{Holland} {et~al.}(2002){Holland}, {Soszy{\' n}ski}, {Gladders},
  {Barrientos}, {Berlind}, {Bersier}, {Garnavich}, {Jha}, \&
  {Stanek}}]{2002AJ....124..639H}
{Holland}, S.~T., {Soszy{\' n}ski}, I., {Gladders}, M.~D., {et~al.} 2002, AJ,
  124, 639

\bibitem[{{In't Zand} {et~al.}(2002){In't Zand}, {Kuiper}, {Heise}, {Piro}, \&
  {Gandolfi}}]{2002GCN..1348....1I}
{In't Zand}, J.~J.~M., {Kuiper}, L., {Heise}, J., {Piro}, L., \& {Gandolfi}, G.
  2002, GCN Circ., 1348

\bibitem[{{Jha} {et~al.}(2001){Jha}, {Pahre}, {Garnavich}, {Calkins},
  {Kilgard}, {Matheson}, {McDowell}, {Roll}, \& {Stanek}}]{2001ApJ...554L.155J}
{Jha}, S., {Pahre}, M.~A., {Garnavich}, P.~M., {et~al.} 2001, ApJ, 554, L155

\bibitem[{{Kouveliotou} {et~al.}(2004){Kouveliotou}, {Woosley}, {Patel},
  {Levan}, {Blandford}, {Ramirez-Ruiz}, {Wijers}, {Weisskopf}, {Tennant},
  {Pian}, \& {Giommi}}]{2004ApJ...608..872K}
{Kouveliotou}, C., {Woosley}, S.~E., {Patel}, S.~K., {et~al.} 2004, ApJ, 608,
  872

\bibitem[{{Le Floc'h} {et~al.}(2003){Le Floc'h}, {Duc}, {Mirabel}, {Sanders},
  {Bosch}, {Diaz}, {Donzelli}, {Rodrigues}, {Courvoisier}, {Greiner},
  {Mereghetti}, {Melnick}, {Maza}, \& {Minniti}}]{2003A&A...400..499L}
{Le Floc'h}, E., {Duc}, P.-A., {Mirabel}, I.~F., {et~al.} 2003, A\&A, 400, 499

\bibitem[{{Levan} {et~al.}(2004){Levan}, {Fruchter}, {Rhoads}, {Mobasher},
  {Tanvir}, {Gorosabel}, {Rol}, \& {Dell'Antonio}}]{Levan:2004}
{Levan}, A., {Fruchter}, A., {Rhoads}, J., {et~al.} 2004, in preparation

\bibitem[{{MacFadyen} \& {Woosley}(1999)}]{1999ApJ...524..262M}
{MacFadyen}, A.~I. \& {Woosley}, S.~E. 1999, ApJ, 524, 262

\bibitem[{{Malesani} {et~al.}(2004){Malesani}, {Tagliaferri}, {Chincarini},
  {Covino}, {Della Valle}, {Fugazza}, {Mazzali}, {Zerbi}, {D'Avanzo},
  {Kalogerakos}, {Simoncelli}, {Antonelli}, {Burderi}, {Campana}, {Cucchiara},
  {Fiore}, {Ghirlanda}, {Goldoni}, {G{\" o}tz}, {Mereghetti}, {Mirabel},
  {Romano}, {Stella}, {Minezaki}, {Yoshii}, \& {Nomoto}}]{2004ApJ...609L...5M}
{Malesani}, D., {Tagliaferri}, G., {Chincarini}, G., {et~al.} 2004, ApJ, 609,
  L5

\bibitem[{{Martini} {et~al.}(2003){Martini}, {Garnavich}, \&
  {Stanek}}]{2003GCN..1980....1M}
{Martini}, P., {Garnavich}, P., \& {Stanek}, K.~Z. 2003, GCN Circ., 1980

\bibitem[{{Masetti} {et~al.}(2001){Masetti}, {Palazzi}, {Pian}, {Mannucci},
  {Antonelli}, {Di Paola}, {Saracco}, {Savaglio}, {Amati}, {Bartolini},
  {Bernabei}, {Bettoni}, {Covino}, {Cristiani}, {Desidera}, {Di Serego
  Alighieri}, {Falomo}, {Frontera}, {Ghinassi}, {Guarnieri}, {Magazz{\ u}},
  {Maiolino}, {Mignoli}, {Nicastro}, {Pedani}, {Piccioni}, {Poggianti},
  {Testa}, {Valentini}, \& {Zacchei}}]{2001A&A...374..382M}
{Masetti}, N., {Palazzi}, E., {Pian}, E., {et~al.} 2001, A\&A, 374, 382393

\bibitem[{{Masetti} {et~al.}(2003){Masetti}, {Palazzi}, {Pian}, {Simoncelli},
  {Hunt}, {Maiorano}, {Levan}, {Christensen}, {Rol}, {Savaglio}, {Falomo},
  {Castro-Tirado}, {Hjorth}, {Delsanti}, {Pannella}, {Mohan}, {Pandey},
  {Sagar}, {Amati}, {Burud}, {Castro Cer{\' o}n}, {Frontera}, {Fruchter},
  {Fynbo}, {Gorosabel}, {Kaper}, {Klose}, {Kouveliotou}, {Nicastro},
  {Pedersen}, {Rhoads}, {Salamanca}, {Tanvir}, {Vreeswijk}, {Wijers}, \& {van
  den Heuvel}}]{2003A&A...404..465M}
{Masetti}, N., {Palazzi}, E., {Pian}, E., {et~al.} 2003, A\&A, 404, 465

\bibitem[{{Matheson} {et~al.}(2003{\natexlab{a}}){Matheson}, {Garnavich},
  {Foltz}, {West}, {Williams}, {Falco}, {Calkins}, {Castander}, {Gawiser},
  {Jha}, {Bersier}, \& {Stanek}}]{2003ApJ...582L...5M}
{Matheson}, T., {Garnavich}, P.~M., {Foltz}, C., {et~al.} 2003{\natexlab{a}},
  ApJ, 582, L5

\bibitem[{{Matheson} {et~al.}(2003{\natexlab{b}}){Matheson}, {Garnavich},
  {Stanek}, {Bersier}, {Holland}, {Krisciunas}, {Caldwell}, {Berlind}, {Bloom},
  {Bolte}, {Bonanos}, {Brown}, {Brown}, {Calkins}, {Challis}, {Chornock},
  {Echevarria}, {Eisenstein}, {Everett}, {Filippenko}, {Flint}, {Foley},
  {Freedman}, {Hamuy}, {Harding}, {Hathi}, {Hicken}, {Hoopes}, {Impey},
  {Jannuzi}, {Jansen}, {Jha}, {Kaluzny}, {Kannappan}, {Kirshner}, {Latham},
  {Lee}, {Leonard}, {Li}, {Luhman}, {Martini}, {Mathis}, {Maza}, {Megeath},
  {Miller}, {Minniti}, {Olszewski}, {Papenkova}, {Phillips}, {Pindor},
  {Sasselov}, {Schild}, {Schweiker}, {Spahr}, {Thomas-Osip}, {Thompson},
  {Weisz}, {Windhorst}, \& {Zaritsky}}]{2003ApJ...599..394M}
{Matheson}, T., {Garnavich}, P.~M., {Stanek}, K.~Z., {et~al.}
  2003{\natexlab{b}}, ApJ, 599, 394

\bibitem[{{Metcalfe} {et~al.}(2003){Metcalfe}, {Kneib}, {McBreen}, {Altieri},
  {Biviano}, {Delaney}, {Elbaz}, {Kessler}, {Leech}, {Okumura}, {Ott},
  {Perez-Martinez}, {Sanchez-Fernandez}, \& {Schulz}}]{2003A&A...407..791M}
{Metcalfe}, L., {Kneib}, J.-P., {McBreen}, B., {et~al.} 2003, A\&A, 407, 791

\bibitem[{{Mirabal} {et~al.}(2003){Mirabal}, {Paerels}, \&
  {Halpern}}]{2003ApJ...587..128M}
{Mirabal}, N., {Paerels}, F., \& {Halpern}, J.~P. 2003, ApJ, 587, 128

\bibitem[{{M{\o}ller} {et~al.}(2002){M{\o}ller}, {Fynbo}, {Hjorth}, {Thomsen},
  {Egholm}, {Andersen}, {Gorosabel}, {Holland}, {Jakobsson}, {Jensen},
  {Pedersen}, {Pedersen}, \& {Weidinger}}]{2002A&A...396L..21M}
{M{\o}ller}, P., {Fynbo}, J.~P.~U., {Hjorth}, J., {et~al.} 2002, A\&A, 396, L21

\bibitem[{{Pedersen} {et~al.}(2003){Pedersen}, {Fynbo}, {Hjorth}, \&
  {Watson}}]{2003GCN..1924....1P}
{Pedersen}, K., {Fynbo}, J., {Hjorth}, J., \& {Watson}, D. 2003, GCN Circ.,
  1924

\bibitem[{{Pedersen} {et~al.}(2004){Pedersen}, {Hurley}, {Hjorth}, {Smith},
  {Andersen}, {Christensen}, {Cline}, \& {Fynbo}}]{Pedersen:2004}
{Pedersen}, K., {Hurley}, K., {Hjorth}, J., {et~al.} 2004, {ApJ submitted}

\bibitem[{{Persic} {et~al.}(2004){Persic}, {Rephaeli}, {Braito}, {Cappi},
  {Della Ceca}, {Franceschini}, \& {Gruber}}]{2004A&A...419..849P}
{Persic}, M., {Rephaeli}, Y., {Braito}, V., {et~al.} 2004, A\&A, 419, 849

\bibitem[{{Piro} {et~al.}(2002){Piro}, {Frail}, {Gorosabel}, {Garmire},
  {Soffitta}, {Amati}, {Andersen}, {Antonelli}, {Berger}, {Frontera}, {Fynbo},
  {Gandolfi}, {Garcia}, {Hjorth}, {Zand}, {Jensen}, {Masetti}, {M{\o}ller},
  {Pedersen}, {Pian}, \& {Wieringa}}]{2002ApJ...577..680P}
{Piro}, L., {Frail}, D.~A., {Gorosabel}, J., {et~al.} 2002, ApJ, 577, 680

\bibitem[{{Piro} {et~al.}(2000){Piro}, {Garmire}, {Garcia}, {Stratta}, {Costa},
  {Feroci}, {M{\' e}sz{\' a}ros}, {Vietri}, {Bradt}, {Frail}, {Frontera},
  {Halpern}, {Heise}, {Hurley}, {Kawai}, {Kippen}, {Marshall}, {Murakami},
  {Sokolov}, {Takeshima}, \& {Yoshida}}]{2000Sci...290..955P}
{Piro}, L., {Garmire}, G., {Garcia}, M., {et~al.} 2000, Sci, 290, 955

\bibitem[{{Price} {et~al.}(2003){Price}, {Kulkarni}, {Berger}, {Fox}, {Bloom},
  {Djorgovski}, {Frail}, {Galama}, {Harrison}, {McCarthy}, {Reichart}, {Sari},
  {Yost}, {Jerjen}, {Flint}, {Phillips}, {Warren}, {Axelrod}, {Chevalier},
  {Holtzman}, {Kimble}, {Schmidt}, {Wheeler}, {Frontera}, {Costa}, {Piro},
  {Hurley}, {Cline}, {Guidorzi}, {Montanari}, {Mazets}, {Golenetskii},
  {Mitrofanov}, {Anfimov}, {Kozyrev}, {Litvak}, {Sanin}, {Boynton}, {Fellows},
  {Harshman}, {Shinohara}, {Gal-Yam}, {Ofek}, \&
  {Lipkin}}]{2003ApJ...589..838P}
{Price}, P.~A., {Kulkarni}, S.~R., {Berger}, E., {et~al.} 2003, ApJ, 589, 838

\bibitem[{{Prochaska} {et~al.}(2004){Prochaska}, {Bloom}, {Chen}, {Hurley},
  {Melbourne}, {Dressler}, {Graham}, {Osip}, \& {Vacca}}]{2004ApJ...611..200P}
{Prochaska}, J.~X., {Bloom}, J.~S., {Chen}, H., {et~al.} 2004, ApJ, 611, 200

\bibitem[{{Ranalli} {et~al.}(2003){Ranalli}, {Comastri}, \&
  {Setti}}]{2003A&A...399...39R}
{Ranalli}, P., {Comastri}, A., \& {Setti}, G. 2003, A\&A, 399, 39

\bibitem[{{Reeves} \& {Watson}(2004)}]{Reeves:2004}
{Reeves}, J.~N. \& {Watson}, D. 2004, in preparation

\bibitem[{{Reeves} {et~al.}(2002){Reeves}, {Watson}, {Osborne}, {Pounds},
  {O'Brien}, {Short}, {Turner}, {Watson}, {Mason}, {Ehle}, \&
  {Schartel}}]{2002Natur.416..512R}
{Reeves}, J.~N., {Watson}, D., {Osborne}, J.~P., {et~al.} 2002, Nat, 416, 512

\bibitem[{{Rol} {et~al.}(2003){Rol}, {Vreeswijk}, \&
  {Jaunsen}}]{2003GCN..1981....1R}
{Rol}, E., {Vreeswijk}, P., \& {Jaunsen}, A. 2003, GCN Circ., 1981

\bibitem[{{Sako} \& {Harrison}(2002)}]{2002GCN..1716....1S}
{Sako}, M. \& {Harrison}, F. 2002, GCN Circ., 1716

\bibitem[{{Sato} {et~al.}(2004){Sato}, {Cowie}, {Kawara}, {Matsuhara}, {Okuda},
  {Sanders}, {Sofue}, {Taniguchi}, \& {Wakamatsu}}]{2004AJ....127.1285S}
{Sato}, Y., {Cowie}, L.~L., {Kawara}, K., {et~al.} 2004, AJ, 127, 1285

\bibitem[{{Schaefer} {et~al.}(2003){Schaefer}, {Gerardy}, {H{\" o}flich},
  {Panaitescu}, {Quimby}, {Mader}, {Hill}, {Kumar}, {Wheeler}, {Eracleous},
  {Sigurdsson}, {M{\' e}sz{\' a}ros}, {Zhang}, {Wang}, {Hessman}, \&
  {Petrosian}}]{2003ApJ...588..387S}
{Schaefer}, B.~E., {Gerardy}, C.~L., {H{\" o}flich}, P., {et~al.} 2003, ApJ,
  588, 387

\bibitem[{{Smith} {et~al.}(1999){Smith}, {Tilanus}, {van Paradijs}, {Galama},
  {Groot}, {Vreeswijk}, {Kouveliotou}, {Wijers}, \&
  {Tanvir}}]{1999A&A...347...92S}
{Smith}, I.~A., {Tilanus}, R.~P.~J., {van Paradijs}, J., {et~al.} 1999, A\&A,
  347, 92

\bibitem[{{Smith} {et~al.}(2001){Smith}, {Tilanus}, {Wijers}, {Tanvir},
  {Vreeswijk}, {Rol}, \& {Kouveliotou}}]{2001A&A...380...81S}
{Smith}, I.~A., {Tilanus}, R.~P.~J., {Wijers}, R.~A.~M.~J., {et~al.} 2001,
  A\&A, 380, 81

\bibitem[{{Soderberg} {et~al.}(2004){Soderberg}, {Kulkarni}, {Berger}, {Fox},
  {Sako}, {Frail}, {Gal-Yam}, {Moon}, {Cenko}, {Yost}, {Phillips}, {Persson},
  {Freedman}, {Wyatt}, {Jayawardhana}, \& {Paulson}}]{2004Natur.430..648S}
{Soderberg}, A.~M., {Kulkarni}, S.~R., {Berger}, E., {et~al.} 2004, Nat, 430,
  648

\bibitem[{{Stanek} {et~al.}(2003){Stanek}, {Matheson}, {Garnavich}, {Martini},
  {Berlind}, {Caldwell}, {Challis}, {Brown}, {Schild}, {Krisciunas}, {Calkins},
  {Lee}, {Hathi}, {Jansen}, {Windhorst}, {Echevarria}, {Eisenstein}, {Pindor},
  {Olszewski}, {Harding}, {Holland}, \& {Bersier}}]{2003ApJ...591L..17S}
{Stanek}, K.~Z., {Matheson}, T., {Garnavich}, P.~M., {et~al.} 2003, ApJ, 591,
  L17

\bibitem[{{Tanvir} {et~al.}(2004){Tanvir}, {Barnard}, {Blain}, {Fruchter},
  {Kouveliotou}, {Natarajan}, {Ramirez-Ruiz}, {Rol}, {Smith}, {Tilanus}, \&
  {Wijers}}]{2004MNRAS.352.1073T}
{Tanvir}, N.~R., {Barnard}, V.~E., {Blain}, A.~W., {et~al.} 2004, MNRAS, 352,
  1073

\bibitem[{{Tiengo} {et~al.}(2004{\natexlab{a}}){Tiengo}, {Mereghetti}, \& {de
  Luca}}]{2004GCN..2548....1T}
{Tiengo}, A., {Mereghetti}, S., \& {de Luca}, A. 2004{\natexlab{a}}, GCN Circ.,
  2548

\bibitem[{{Tiengo} {et~al.}(2004{\natexlab{b}}){Tiengo}, {Mereghetti},
  {Ghisellini}, {Tavecchio}, \& {Ghirlanda}}]{2004A&A...423..861T}
{Tiengo}, A., {Mereghetti}, S., {Ghisellini}, G., {Tavecchio}, F., \&
  {Ghirlanda}, G. 2004{\natexlab{b}}, A\&A, 423, 861

\bibitem[{{Vanderspek} {et~al.}(2002){Vanderspek}, {Marshall}, {Ford}, \&
  {Ricker}}]{2002GCN..1504....1V}
{Vanderspek}, R., {Marshall}, H.~L., {Ford}, P.~G., \& {Ricker}, G.~R. 2002,
  GCN Circ., 1504

\bibitem[{{Vreeswijk} {et~al.}(1999){Vreeswijk}, {Rol}, {Hjorth},
  {Kouveliotou}, {Pian}, {Palazzi}, {Pedersen}, {Gorosabel}, {Castro-Tirado},
  \& {Greiner}}]{1999GCN...496....1V}
{Vreeswijk}, P.~M., {Rol}, E., {Hjorth}, J., {et~al.} 1999, GCN Circ., 496

\bibitem[{{Watson} {et~al.}(2003){Watson}, {Reeves}, {Hjorth}, {Jakobsson}, \&
  {Pedersen}}]{2003ApJ...595L..29W}
{Watson}, D., {Reeves}, J.~N., {Hjorth}, J., {Jakobsson}, P., \& {Pedersen}, K.
  2003, ApJ, 595, L29

\bibitem[{{Watson} {et~al.}(2002{\natexlab{a}}){Watson}, {Reeves}, {Osborne},
  {O'Brien}, {Pounds}, {Tedds}, {Santos-Ll\'eo}, \&
  {Ehle}}]{2002A&A...393L...1W}
{Watson}, D., {Reeves}, J.~N., {Osborne}, J., {et~al.} 2002{\natexlab{a}},
  A\&A, 393, L1

\bibitem[{{Watson} {et~al.}(2002{\natexlab{b}}){Watson}, {Reeves}, {Osborne},
  {Tedds}, {O'Brien}, {Tomas}, \& {Ehle}}]{2002A&A...395L..41W}
{Watson}, D., {Reeves}, J.~N., {Osborne}, J.~P., {et~al.} 2002{\natexlab{b}},
  A\&A, 395, L41

\end{thebibliography}

\end{document}